# Dismantling a dogma: the inflated significance of neutral genetic diversity in conservation genetics

João C. Teixeira[1,2*] and Christian D. Huber[1*]

[1]School of Biological Sciences, The University of Adelaide, 5005 Adelaide, Australia

[2]ARC Centre of Excellence for Australian Biodiversity and Heritage (CABAH), The University of Adelaide, 5005 Adelaide, Australia

*Corresponding authors. E-mail: joao.teixeira@adelaide.edu.au; christian.huber@adelaide.edu.au;

**Abstract**

The current rate of species extinction is rapidly approaching unprecedented highs and life on Earth presently faces a sixth mass extinction event driven by anthropogenic activity, climate change and ecological collapse. The field of conservation genetics aims at preserving species by using their levels of genetic diversity, usually measured as neutral genome-wide diversity, as a barometer for evaluating population health and extinction risk. A fundamental assumption is that higher levels of genetic diversity lead to an increase in fitness and long-term survival of a species. Here, we argue against the perceived importance of neutral genetic diversity for the conservation of wild populations and species. We demonstrate that no simple general relationship exists between neutral genetic diversity and the risk of species extinction. Instead, a better understanding of the properties of functional genetic diversity, demographic history, and ecological relationships, is necessary for developing and implementing effective conservation genetic strategies.

**Are species with little genetic diversity endangered?**

Climate change caused by human activity is currently responsible for widespread ecological disruption and habitat destruction, with an ensuing unprecedented rate of species loss known as the Anthropocene Mass Extinction[1–4]. This catastrophic scenario poses a serious threat to the future of life and human survival on Earth and has sparked a global sense of emergency about the need to preserve the diversity of life on the planet. However, this emergency has also fostered the development and implementation of imperfect and pragmatic conservation strategies with potential detrimental consequences for the preservation of life on Earth[5].

A fundamental conception underlying many of these strategies is the importance attributed to intraspecific genetic diversity, measured at markers scattered across the genome, for assessing the extinction risk of species facing rapid environmental change[6–8]. Specifically, low genetic diversity is often interpreted as an indicator for inbreeding depression and increased genetic drift, and related to reduced individual lifespan, health and survival, along with a depleted capacity for population growth[9]. In contrast, high levels of genetic diversity are seen as key to promote population survival and guaranteeing the adaptive potential of natural populations in the face of rapidly changing environmental pressures[10]. These principles are reflected in strategies such as genetic rescue, where the genetic diversity of a threatened or endangered population is increased by facilitating gene-flow from a population with high levels of diversity[11].

However, supporting empirical evidence for the existence of a causal relationship between genetic diversity and population fitness or adaptive potential is weak. While there are some examples of endangered species with low levels of genetic diversity, they do not necessarily constitute the norm. For example, the Sumatran orangutan (*Pongo abelii*), one of our closest living relatives, has one of the largest effective population sizes among the great apes, with levels of neutral genetic diversity almost twice that of humans[12]. Nonetheless, the species is currently at a high risk of extinction as a result of deforestation, and its conservation status has been listed as *critically endangered* by the International Union for the Conservation of Nature (IUCN). Another example is that of the Australian koala (*Phascolarctos cinereus*), which has recently been regarded as a '*complex conservation conundrum*'[13]. The southern populations of koalas show very low levels of genetic diversity as a result of serial bottlenecks, but are so numerous that starvation problems have been reported across the state of South Australia[14]. At the same time, the genetically richer koala populations from New South Wales and Queensland are severely under threat[13].

If genetic diversity is indeed a major factor affecting the health and survival of populations in the wild, then one would expect endangered species to show, on average, lower levels of genomic diversity. However, the IUCN Red List status is only a poor predictor of a species' genome-wide heterozygosity[15–18] (Figure 1). It has been previously argued that this lack of correlation reflects a deficient classification and that genome-wide patterns of neutral diversity should be incorporated into IUCN's listing criteria to more accurately assess the likelihood of future extinction and the associated need for conservation[17,19,20]. However, a more parsimonious explanation is that the lack of correlation is due to genetic diversity having only a minor predictive role for the fitness or extinction risk of a species. This is consistent with heterozygosity-fitness correlation (HFC) studies in animal populations where, on average, only 1% of variance in fitness is explained by levels of heterozygosity[21,22]. Furthermore, the increased rate of genetic drift in small populations inevitably leads to low levels of neutral genetic diversity over time. Hence, it is expected that some endangered species show low levels of neutral genetic diversity as a result of small population numbers and high levels of genetic drift, but this does not imply that genetic factors, such as those listed in Table 1, are causally responsible for decreasing population sizes or have an impact on population health and survival.

**Table 1. Classification of genetic extinction models**

| Extinction model (main cause) | Trigger | Genetic mechanism leading to reduced fitness | Selected references |
|---|---|---|---|
| Inbreeding depression | Breeding of related individuals | Recessive deleterious mutations become homozygous due to inbreeding (although other mechanisms have been proposed) | Charlesworth and Willis, 2009[23]; Hedrick and Garcia-Dorado 2016[24] |
| Mutational meltdown | Ineffective selection due to small/reduced population size | Many slightly deleterious mutations become fixed due to strong genetic drift in small populations | Lynch, Conery and Buerger 1995[25]; Agrawal and Whitlock 2012[26] |
| Maladaptation to environment | Changing environment (but see Brady et al. 2019[27]) | Organisms are maladapted to the environment | Tallmon, Luikart and Waples 2004[28]; Gomulkiewicz and Holdt, 1995[29] |

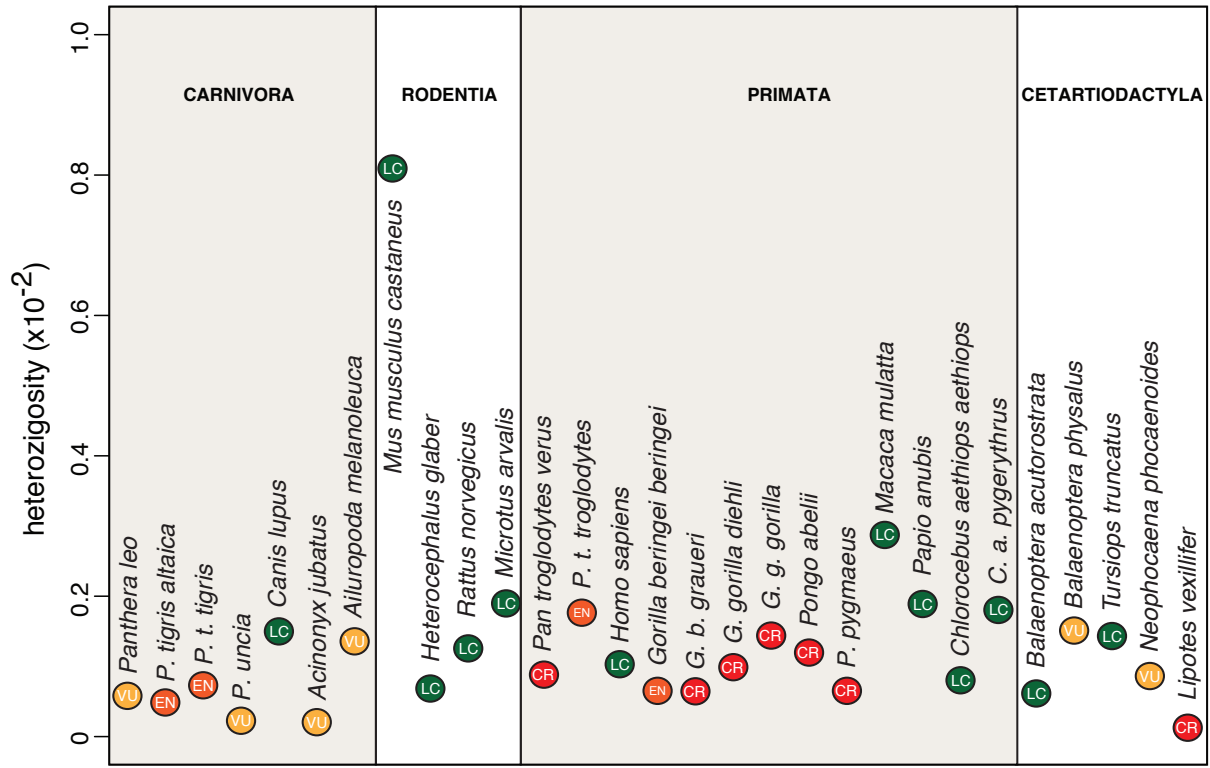

**Figure 1. Genome-wide heterozygosity is a poor predictor of IUCN's Red List status.** Mammalian heterozygosity estimates were taken from Robinson et al.[30]. Some critically endangered (CR) species, such as the Yangtze river dolphin or Baiji (*Lipotes vexillifer*), show extremely low levels of heterozygosity. However, low levels of heterozygosity are not necessarily due to recent anthropogenic pressures (the observed low genetic diversity in the Baiji is due to a population bottleneck during the last glacial maximum ~20kya[31]). Other endangered (EN) and critically endangered species, such as the Sumatran Orangutan (*Pongo abelii*), show levels of heterozygosity that fall well within the range of least concern (LC) species.

There is extensive scientific debate regarding the relative roles that genetics and demography play in the extinction of populations, with many arguments both for and against the significance of genetic factors[9,32–38]. Here, we provide our perspective specifically on the significance of neutral or genome-wide genetic diversity as an indicator of the conservation status of a species, taking into account recent developments in population genetics as well as empirical results that were made possible by the eruption of genomic datasets becoming available in the last decade. We propose that neutral genetic diversity has only very limited relevance for conservation genetics. Instead, individual genomes harbour functionally relevant, genomically localised variation that can severely impact population fitness and should, therefore, guide conservation efforts. Population genetic models suggest that the relationship between neutral diversity, functional diversity, and population fitness, can be un-intuitive and rely on several unknown parameters. Thus, we argue for the necessity of precisely mapping adaptive genetic variation in the genome, and for a better understanding of mutation load and its consequences, as these aspects are paramount for developing and implementing effective conservation genetic strategies in the face of environmental change and the ongoing mass extinction.

**Inbreeding depression and hybrid vigour**

Two processes often cited in the conservation genetics literature to link the amount of neutral genetic diversity to population fitness are *inbreeding depression* and *hybrid vigour*. The impact each of these factors has on fitness arises from the exposure or masking of recessive deleterious variants[23]. In the case of inbreeding depression, the reduction in fitness of a population is caused by mating between related individuals as a result, for example, of habitat fragmentation[39]. Such consanguineous mating leads to an increase in genomic segments of identity by descent harbouring recessive deleterious variants that become exposed in homozygosity in some individuals (Figure 2), and in extreme cases can lead to the extinction of populations or species (Table 1). An inverse process occurs in the case of outbreeding, where a fitness increase in the population can be observed as a result of hybrid vigour, as recessive deleterious mutations become masked in heterozygous states in hybrid individuals[40] (Figure 2).

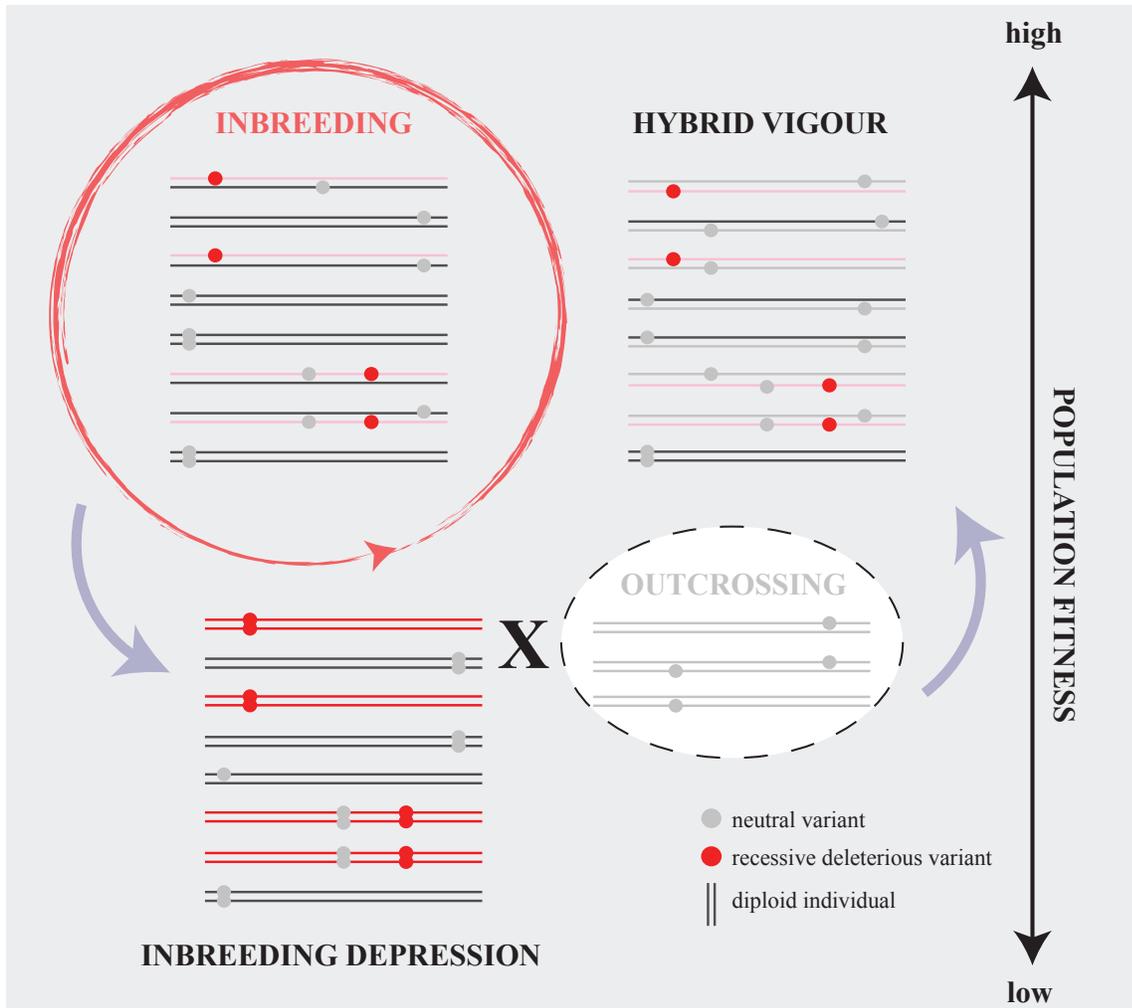

**Figure 2. Inbreeding depression and hybrid vigour.** Inbreeding leads to an increase of homozygous genotypes. Recessive deleterious variants that become homozygous lead to a reduction in fitness in affected individuals (i.e., inbreeding depression). Outcrossing with individuals from a different population reduces homozygous genotypes and thus increases population fitness again (hybrid vigour).

Although low levels of genetic diversity might indicate high rates of consanguinity, there are species with strikingly low levels of diversity but no signs of inbreeding (e.g. the brown hyena[41]). Moreover, inbreeding depression can also occur in high diversity populations[42]. In fact, the relations between inbreeding depression, hybrid vigour and genetic diversity are more complex than often regarded. A recent study by Kyriazis et al. has proposed that using a population with high levels of genetic diversity for genetic rescue can be ineffective for preventing the extinction of small populations[43]. By leveraging population genetic simulations on ecological models, the authors demonstrated that this occurs via the introduction of a large number of recessive deleterious variants whose combined effects are potentially more harmful than the increase in fitness associated with hybrid vigour. Moreover, they showed that outcrossing with a population with low genetic diversity can introduce deleterious variants that had drifted to fixation due to small population sizes[43]. Thus, a population with intermediate levels of genetic diversity is most effective for rescuing a small endangered population. Furthermore, it has been shown that the genetic basis of inbreeding depression can depend on the specific demographic history of the population[44] and that the timing and nature of bottlenecks can shape the levels of genetic load[45].

Considering more complex relationships between genetic variants and fitness, such as epistasis and gene-environment interactions, makes the commonly assumed relationships between inbreeding, outcrossing, and fitness even less predictable. Facilitated outcrossing and migration can lead to a species-wide reduction in diversity and prevent local adaptation, leading to a further long-term negative effect on species survival[46]. Furthermore, as populations diverge and ultimately evolve into distinct species, hybrids between them gradually become inviable and infertile. Thus, even though outcrossing might prevent inbreeding depression, it can also lead to reduced fitness because of non-compatible mutations within hybrid individuals. Although such outbreeding depression is considered rare, a delayed onset of outbreeding depression until F3 and later generations has not been well examined[47]. Further, outcrossing depression between diverged populations is difficult to predict[47], and we are only beginning to understand how hybrid vigour over time transforms into hybrid inviability as a function of genetic drift and the individual and epistatic fitness effects of mutations[48]. A better understanding of population genetic mechanisms will allow us to evaluate the relative risks of inbreeding and outcrossing depression, but it is clear that a simple relation with neutral genetic diversity is not warranted.

**Low genetic diversity and mutation load**

Besides the effects of inbreeding on fitness, it is also often assumed that low genetic diversity indicates that selection is ineffective and random genetic drift dominates allele frequency dynamics. Thus, deleterious mutations are not effectively removed and/or increase in frequency, resulting in reduced population fitness (i.e. mutation load, see Box 1), leading to a vicious circle that includes further reduced population size, even less effective selection and, eventually, the extinction of the population (a process often referred to as "mutational meltdown", see Table 1). Although we do not discount that, over long periods of time, a lack of effective selection and the accumulation of deleterious mutations will most certainly have an effect on the genomic and physiological integrity of an organism, it is not exactly clear when reduced neutral genetic diversity is indicating problematic levels of genetic drift, or when population size becomes too small to prevent a decline in population fitness. Experiments in *Drosophila* suggest that the species can survive for many generations with population sizes as small as 25 individuals without showing any signs of reduced fitness compared to outbred wild populations[49]. Furthermore, certain species, such as the San Nicolas Island Fox[30], the Vaquita Porpoise[50], or the Brown Hyena[41], have a near absence of genetic variation, even though no fitness reduction or any apparent genetically-linked diseases have been observed, which further questions the importance of genetic diversity for long-term population persistence. Hence, a better understanding of how deleterious mutations act in the context of a species' ecological interactions is essential to better understand how mutation load affects its survival (Box 1).

**Box 1: Estimating and interpreting mutation load**

Mutation load is a reduction in fitness of a population caused by the constant pressure of deleterious mutations appearing in the genomes of individuals, which was first theoretically conceptualized by Haldane in 1937[26,51]. Recent advances in genome sequencing technology have allowed us to empirically investigate mutation load and test theoretical predictions using genomic data from humans[52–54], gorillas[55], dogs[56], horses[57], Alpine ibex[18], pandas[58], and several other species[59]. In these studies, the estimation of load relies on identifying deleterious mutations in the genome, which is usually achieved using measures of phylogenetic similarity. If a mutation within an individual appears at a site that is conserved (i.e. is the same nucleotide across many different phylogenetically diverged species) then it is assumed that this mutation has a negative effect on that individual's fitness. For each individual within a population, the total number of mutations at phylogenetically constrained sites can be added up to arrive at a statistic that is proportional to mutation load[60,61]. However, it is important to note that this score makes multiple assumptions: 1) it is possible to accurately identify deleterious mutations and their selection coefficient; 2) deleterious mutations act additively; 3) there is no epistasis. However, available comparative genomic methods do not reliably detect the majority of selected sites[62] and do not differentiate between mutations with minor effects on fitness from those with drastic effects on fitness[62,63]. Furthermore, we do not have a good understanding of the distribution of mutational fitness parameters such as dominance and epistasis[61,64–66]. Thus, it is not yet entirely clear how much interspecies comparisons of mutation load truly reflect underlying differences in population fitness.

Apart from the difficulty of measuring mutation load, arguably the biggest obstacle for understanding its relevance in conservation genetics is the question of what mutation load actually means for the size and survival of a population. Mutation load is a population genetics concept regarding relative fitness to an idealized individual that does not carry any deleterious mutations, i.e. to the perfect genotype. However, it was mathematically shown that the perfect genotype is exceedingly unlikely to exist[67]. Thus, individuals in a population never directly compete with the perfect genotype. Furthermore, the viability of a population in its ecological and environmental context, i.e. its *absolute* fitness, is more relevant for conservation genetic purposes than relative fitness[26,68]. Agrawal and Whitlock have investigated models that try to close the gap between population genetic mechanisms and the ecological consequences of deleterious mutations[26]. In their model, mutation load can affect birth and death rates, and an individual's rate of resource acquisition. Whereas birth and death rates affect the equilibrium population size, the rate of resource acquisition does not. Individuals that can acquire more resources produce more offspring and, in case mutation load reduces the rate of resource acquisition in some individuals, others can gather more resources and produce more offspring, leading to a constant equilibrium population size. Moreover, if mutation load leads to an individual's early death, e.g. in the zygote state, then this individual does not exhaust resources and the population size is, again, not affected[26]. On the other hand, in models where two species compete for the same resources, mutation load on the rate of resource acquisition does indeed have a strong effect on equilibrium population size and can lead to population extinction[26]. Although these models are agnostic to the many complex ecological dependencies between individuals, populations, and species, they emphasise the importance of ecological relationships for gaining an understanding of how mutation load affects the persistence and health of a population - a factor that is almost always ignored in research studying mutation load.

> In short, mutation load does not always affect populations the same way as it does individuals. In fact, deleterious mutations at different loci, even with the same effect on fitness and the same mutation rate, do not necessarily have the same effect at the population level if they affect different fitness components. Thus, it is important not only to understand which mutations in the genome are selected but also when and how mutations act in the ecological context of the species.

The recent availability of genomic datasets has made it possible to directly estimate and compare the mutation load across different species (Box 1). Contrary to expectations commonly assumed in conservation genetics, species with small population census size actually contain *less* (additive) mutation load than species with larger population sizes[59]. Importantly, mutation load is also not related to IUCN's conservation status[59].

Notwithstanding, measuring differences in mutation load between diverged species is challenging and potentially prone to bioinformatic biases and measurement errors[60] (Box 1). A more tractable question is how demographic differences between populations of the same species have affected the distribution of genetic diversity and load. In this case, only mutations that segregate in the ancestral population of the species have to be considered, and the alignment of genomes is simpler than when comparing genomes from different species[60,61]. Results point to an existing but weak relation between neutral diversity and mutation load in bottlenecked populations[53]. For example, in humans, the Out-of-Africa bottleneck ~50,000 years ago has strongly reduced neutral genetic diversity in non-African human populations. However, even in one of the most extremely bottlenecked human populations, the Greenlandic Inuit, the estimated additive mutation load shows, at most, only a slight increase compared to African populations[69]. Although the Inuit carry deleterious mutations at higher frequencies than other populations, they also carry fewer deleterious variants overall. This is consistent with population genetics theory that suggests that the number of deleterious variants per individual is only modestly affected by a reduction in population size, since many deleterious mutations are lost during the bottleneck[53,60,70]. More generally, a significant increase in additive mutation load is expected only under severe and extended reductions in population size that exceed that of the human Out-of-Africa bottleneck[53] and lead to a reduction in fitness as a result of an increase in the frequency of slightly deleterious mutations through drift (Figure 3). Mutation load due to strongly recessive mutations can show a more immediate response to changes in population size or surges of inbreeding[60,69]. However, similar to additive load, the long-term equilibrium value hardly depends on population size and thus can not be predicted from neutral diversity (Figure 3).

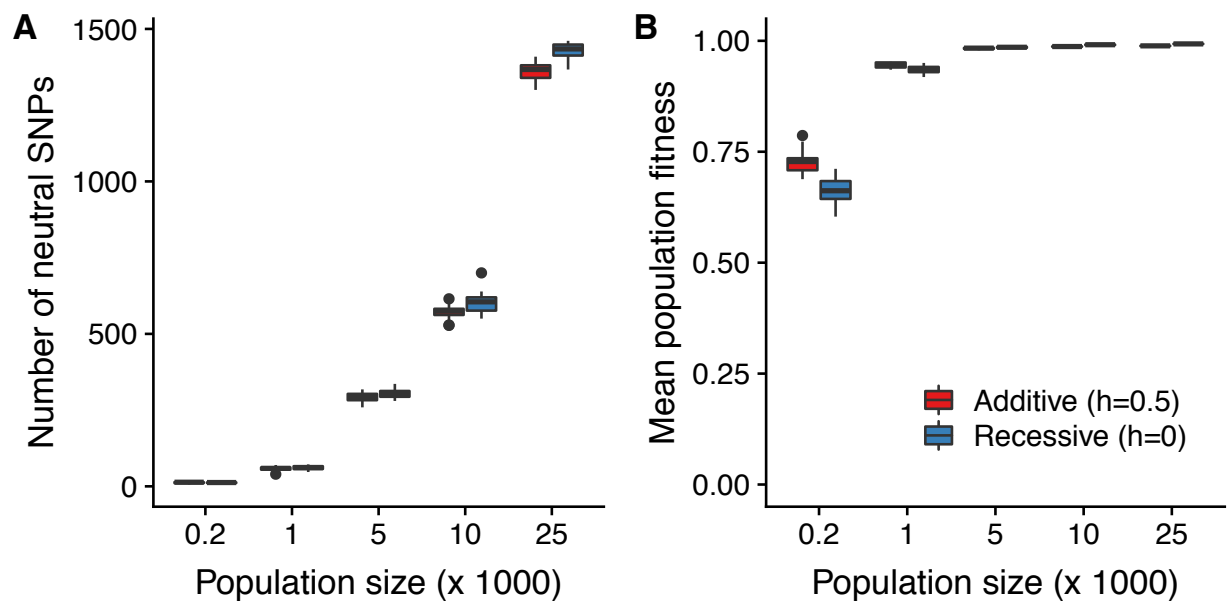

**Figure 3. Forward in time simulations over a wide range of population sizes show mutation load is only modestly affected by $N_e$.** The forward in time population genetic simulations with SLiM 3[71] assumed a constant population size, ranging from 200 to 25,000 individuals. We simulated a total of 1000 evenly distributed exons over

a genetic length of 10 cM, with a nonsynonymous to synonymous ratio of 2:1, and a mutation rate of 1e-5 per exon. The selection coefficient of the nonsynonymous mutations was sampled from a gamma distribution as estimated in Huber et al.[72], and synonymous mutations were assumed to be neutral. We considered either additive ($h$=0.5; red) or recessive ($h$=0; blue) deleterious mutations. We simulated 100,000 generations and recorded genetic diversity and mean fitness of the population at the end of each run. (A) The number of segregating synonymous sites is proportional to the population size, as expected from neutral theory. (B) Above a long-term population size of 5,000 the mean population fitness is close to the maximum, and independent of population size. Only with a fairly small population size of less than 1,000 individuals over prolonged periods of time does fitness become severely reduced due to the fixation of slightly deleterious mutations. More general studies that allow for beneficial mutations and epistasis similarly conclude that only a few hundred individuals are necessary to prevent a decline in fitness via fixation of deleterious mutations[73].

**Neutral diversity does not predict adaptive potential**

Another key concept in the conservation genetic literature is that of adaptive (or evolutionary) potential, which can be defined as the ability of populations to respond to shifts in environmental and selective pressures by means of phenotypic and/or molecular changes[74] and thus prevent the extinction due to maladaptation (Table 1). Accordingly, it is assumed that populations with higher genetic diversity have more adaptive potential because of higher levels of standing genetic variation, which makes them more robust to changing environmental conditions and, thus, more suitable for conservation efforts[75]. However, evidence from surveying genetic diversity of wild populations is showing that such a relationship might not always exist. A study evaluated the effects of introducing steelhead trout (*Oncorhynchus mykiss*) from California into Lake Michigan in the 1880s[76], a species known to hatch in rivers and live in the ocean before returning to freshwater for spawning. Upon its introduction to Lake Michigan, steelhead trout began to use the lake as a surrogate ocean and, despite this remarkable shift in environmental conditions and a strong population bottleneck, showed distinct signatures of local adaptation[76]. Similarly, an experimental study in brook trout (*Salvelinus fontinalis*) suggests that low genetic diversity does not prevent or predict transplantation success into fishless ponds over a wide gradient of ecological variables[77]. As another example, dramatically bottlenecked northern European populations of *Arabidopsis lyrata* and *Arabidopsis thaliana* show strong genomic footprints of adaptation[78,79] and experimental evidence suggests that their strongly reduced genetic diversity has not affected their ability to adapt to local environmental conditions[80–82]. To investigate the determinants of the rate of genetic adaptation more generally, recent studies have analysed genome-wide polymorphism and divergence data from many different species and concluded that low-diversity taxa do not seem to accumulate adaptive substitutions at a substantially slower rate than high-diversity taxa[83,84], although within certain groups such a relationship might exist[84,85].

These results question the existence of a simple general relationship between the levels of genome-wide genetic diversity observed in a population and its potential to genetically adapt to changing environmental conditions. Instead, the nature of genetic diversity segregating at particular *loci* seems to be substantially more important[86,87], and selection acting on these variants makes their distribution and evolutionary trajectories potentially quite distinct from neutral variants. As stated by Lewontin in 1974, '*The question was never really how much genetic variation is there but rather what is the nature of genetic variation for fitness in a population*'[88]. Genetic diversity at few selected *loci* can be maintained for millions of years in species or populations via balancing selection[89,90], including overdominance or heterozygote advantage, temporal/spatial variation in selective pressures, negative frequency-dependent selection, and pleiotropy[91]. In fact, a large fraction of variability in fitness appears to reflect some form of balancing selection[92], and this might constitute a widespread and important mechanism that enables rapid adaptation[93]. The classical example of balancing selection is the Major Histocompatibility Complex (MHC) region, which is responsible for antigen presentation and immune response in vertebrate species[91]. Interestingly, when examined closely, some notorious examples of conservation genetic studies show that a correlation between genetic diversity and adaptive potential is due to polymorphisms segregating at a few *loci* (including the MHC region) where advantageous diversity is maintained through balancing selection. A recent study on bottlenose dolphin populations (*Tursiops aduncus*) argues for an evaluation of the levels of diversity contained within the MHC region, rather genome-wide patterns of neutral genetic diversity, for conservation purposes[94]. Similarly, in a study involving invasive cane toads (*Rhinella marina*) in Australia, the levels of genetic diversity at *loci* involved in resistance to heat and dehydration were either weakly or not at all correlated with effective population size, even in the case of severe bottlenecks[95]. Notably, MHC diversity is not related to effective population size when comparing central chimpanzees and humans[96], but is strongly correlated with environmental factors, such as pathogen load, in the latter[97].

Balancing selection is not the only mechanism distorting a potential relation between neutral diversity and adaptive potential. If adaptation involves pre-existing mutations that shift from effectively deleterious to beneficial as the environment changes, then adaptation can be almost unaffected by population size since the frequency of segregating deleterious mutations is likely to be higher in small populations due to less effective selection (Figure 4). Evidence from experimental evolution and empirical population genetic studies indicates that beneficial mutations are typically deleterious in the absence of selective pressures, suggesting that such dynamics likely occur in nature. Similarly, quantitative genetic models show that stabilizing selection and pleiotropy can strongly reduce the correlation between neutral diversity and additive variation of a quantitative trait[75]. Importantly, neutral genetic diversity is determined by demographic events occurring in the history of a population, including ancient bottlenecks and migration events, whereas the emergence and fate of beneficial mutations after environmental change depends on the population size immediately after the change[98,99] (see also Figure 4). Hence, in order to better predict the adaptive potential of populations, it is necessary to estimate the immediate effective population size, the effect size and rate of mutations that contribute to adaptation, and the mutational target size[99,100], all of which are parameters that are difficult to infer[101].

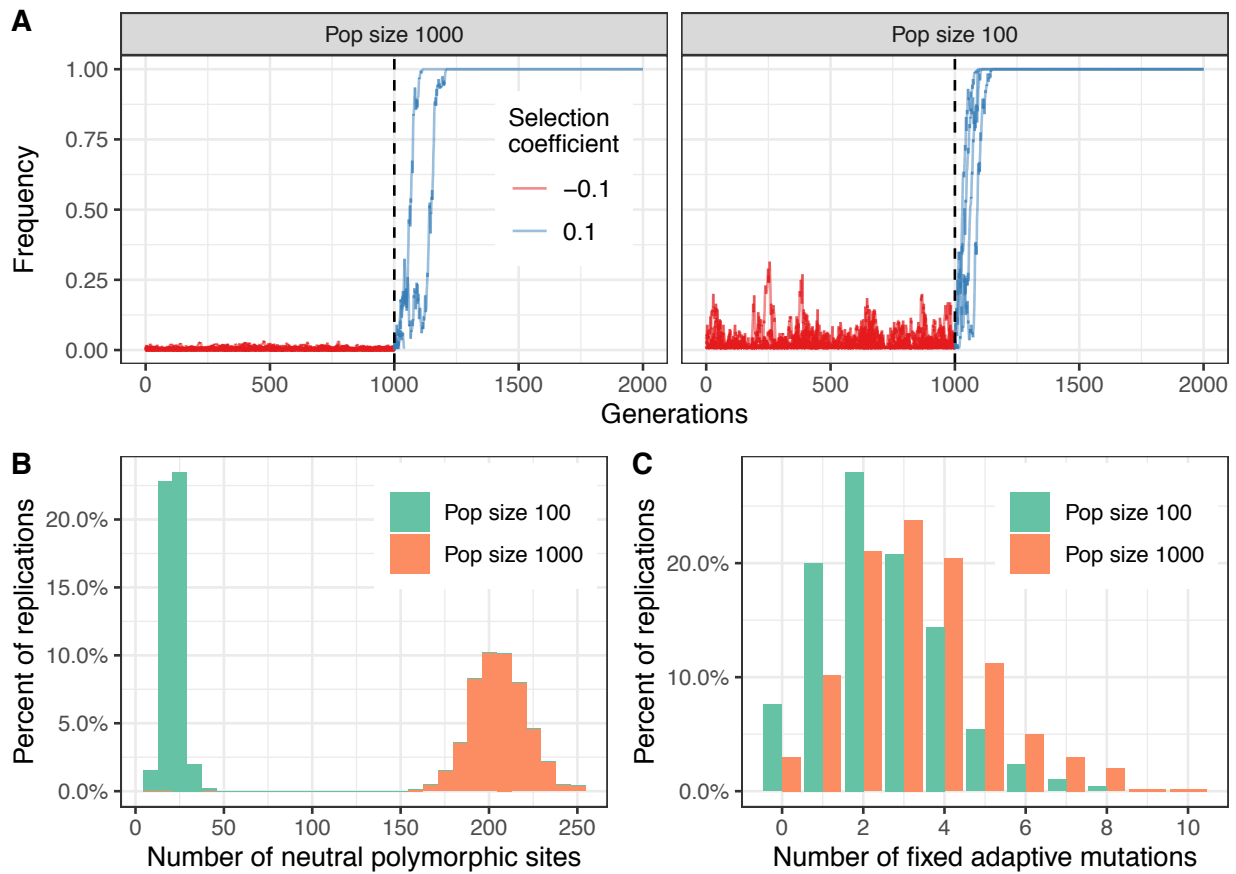

**Figure 4. Adaptive potential can be fairly unrelated to neutral genetic diversity.** In this example, adaptation involves pre-existing mutations that shift from deleterious (s = -0.1) to beneficial (s = 0.1) to mimic the response of two different populations to a shift in environmental conditions. Here, selection is strong enough such that deleterious mutations are effectively selected and thus unlikely to fix even in the smaller population, which is a fundamental assumption of this model. (A) The allele frequency trajectories of selected mutations are plotted as a function of time (note that the trajectories of multiple mutations are overlapping). Before the change in environmental conditions at generation 1,000 (dashed vertical line), the two populations had vastly different population sizes of 1,000 (left) and 100 individuals (right), respectively. In the larger population, where selection is more effective, deleterious mutations segregate at lower frequencies (red lines, left) compared to the smaller population (red lines, right). After the change in environmental conditions, both populations are comprised of 100 individuals, and segregating deleterious mutations become beneficial (blue lines). (B) Neutral genetic diversity right after the change in environmental conditions is, as expected, ten times larger in the historically larger population. (C) However, because the fixation probability increases with allele frequency and deleterious mutations segregate at higher frequencies in smaller populations, the number of fixed adaptive mutations after the change in environmental conditions has a very similar distribution in both populations, whereby the effects of historical population size are much less apparent. The plots in (B) and (C) summarize results for 500 replicates.

Finally, it is often argued that genetic rescue increases the genetic diversity and, therefore, the adaptive potential of low-diversity populations. This implies that an increase in fitness is not predominantly caused by heterosis, but rather that adaptive mutations are transferred from the large to the small population, i.e. there is adaptive variation segregating at high enough frequencies in the source (larger) population. Such an assumption depends, as above, on the mutation rate, the mutational target size, and the effective population size[99–101], but also on past environmental conditions that might have shaped adaptive diversity in the source population. This is nicely exemplified by adaptive introgression that resulted from a series of admixture events between archaic and modern humans[102–110]. Even though Neanderthals and Denisovans had very low levels of genetic diversity and small long-term effective population sizes, the two groups were exposed to harsh climatic and biotic conditions for thousands of generations and thus were able to accumulate adaptive genetic variants that were then introgressed into modern human populations and facilitated adaptation to high altitude[104], pathogens[109] and cold climatic conditions[107].

Accordingly, the choice to concentrate conservation efforts on populations with higher levels of genetic diversity might only be effective in a very restricted set of scenarios, which argues for more informed population genetic approaches to genetic rescue[111,112].

**The future of conservation genetics in the face of mass extinction**

We provide several empirical and theoretical arguments that challenge the commonly assumed relation between genome-wide patterns of neutral genetic diversity and population fitness or adaptive potential. We argue that such a simplistic relation remains speculative and should thus be excluded from conservation strategies. Instead, we urge that a detailed understanding of the genetic basis of deleterious and beneficial variation and an implementation of evolutionary approaches are paramount for the effective conservation of natural populations currently facing unprecedentedly rapid environmental changes.

Specifically, we advocate for a better understanding of crucial parameters, including *dominance* and *epistasis*, to be able to predict the effects of reduced population size, inbreeding, and outcrossing, on fitness and adaptive potential[54,60,61,64,65,72,75,113]. While the estimation of these parameters remains quite challenging, recent work has shown novel ways of inference using model-based approaches[48,56,66,114,115]. Moreover, it is essential to understand how past demography affects the distribution of additive and recessive deleterious variation in order to avoid mutation load and inbreeding depression. We argue for adopting model-based approaches that take into account non-equilibrium demography for understanding the distribution of functional diversity and population fitness[116], and for a shift in conservation genetic principles from a simplistic attempt to increase genetic diversity in wild populations to minimizing genetic load by explicitly considering the fitness effects of deleterious mutations in functional regions of the genome[43]. Bioinformatic and comparative genomic methods can provide valuable information about the location of deleterious variation[117], but their reliance on assumptions should be carefully evaluated[62].

Furthermore, we argue for the need to better understand the genetic basis of adaptation regarding the underlying mode of selection and the genetic architecture behind population responses to changing environmental conditions (e.g. if selection is predominantly acting on standing variation or new mutations)[100,101]. Similarly, it is imperative that conservation efforts concentrate on the nature of genetic diversity at specific functional *loci* in threatened populations, e.g. regions where diversity is maintained by balancing selection, such as the MHC, and that can survive severe bottlenecks and lead to adaptation to changing conditions[118]. Hence, it is key to identify relevant genes and pathways that contain signatures of adaptation by screening whole-genome data in wild populations[119], while controlling for specific demographic events that might distort past signatures of selection (e.g. due to admixture[120]). Recent empirical and experimental studies illustrate the exciting potential for mapping adaptive variation and incorporating genomics to study and predict traits of conservation importance[121–123].

Finally, we agree that the best way to preserve biodiversity is to restore and conserve natural ecosystems, as has been previously proposed[77]. The protection or re-establishment of a functioning ecosystem might provide a more effective alternative to treating single species as isolated, independently evolving entities[124–127].

**Acknowledgments**

We thank Chris Kyriazis, Athanasios Kousathanas, Alexander Salis, and Joshua Schmidt for helpful comments on the manuscript. JCT and CDH are supported by the Australian Research Council through Discovery Grants IN180100017 (JCT) and DP190103606 (CDH), and an ARC DECRA Fellowship DE180100883 (CDH).